# Statistical analysis of C and S Main Belt Asteroids

A. Carbognani

*Astronomical Observatory of the Autonomous Region of the Aosta Valley (OAVdA)*
*Saint-Barthélemy Loc. Lignan, 39*
*11020 Nus (AO) – Italy*

Number of manuscript pages: 22

Number of figures: 8

Number of tables: 2

Editorial correspondence to:

Dr Albino Carbognani

Astronomical Observatory of the Autonomous Region of the Aosta Valley (OAVdA)

Fondazione Clément Fillietroz-ONLUS

Saint-Barthélemy Loc. Lignan, 39

11020 Nus (AO) – Italy

Telephone: +39 0165 77 00 50

Fax: +39 0165 77 00 51

E-mail address: albino.carbognani@alice.it

**Abstract**

In this paper we compare the observable properties of 962 numbered MBAs (Main Belt Asteroids) of Tholen/SMASSII C and S class, with diameter in the range 1-500 km, not belonging to families or binary systems. Above 20 km, the diameters distributions of C and S are similar while under 20 km there is a clear observative bias in favour of small S asteroids which prevents a direct comparison. There is a significant correlation between rotation frequency and diameter both for C and S: if the diameter decreases the rotation frequency tends to increase. There is also a significant correlation between the lightcurve amplitude and the diameter for both samples: if the diameter decreases the lightcurve amplitude tends to increase. For larger diameter the C amplitude tends to be systematically higher than S amplitude of about 0.1 magnitude, but the difference is not very significant. Between 48 and 200 km, the C asteroids have a rotation frequency distribution compatible with a Maxwellian. On the other side, for S asteroids, the compatibility with the Maxwellian concernes diameters greater than 33 km. Considering the rotational properties and the lightcurve amplitude it appears that there are no substantial differences between the samples of C and S asteroids taken into account, and this indicates a good homogeneity in the processes of collisional evolution.

# 1 Introduction

From the observative point of view, the physical quantities easier to be measured on asteroids are the rotation period and the lightcurve amplitude. Usually, it is sufficient to make differential photometry for some nights to obtain these data with reasonable precision (Pravec et al., 2002). For this reason, the data of the rotational periods and the lightcurve amplitude grow more quickly than others. In this paper we compare the rotational properties of numbered MBAs (Main Belt Asteroids) belonging to the Tholen/SMASSII C and S classes (Tholen, 1989, Bus and Binzel, 2002), with dimensions between 1 and 500 km, not belonging to families or binary systems and with a U quality of the lightcurve equal or higher than 2. A U value equal to 2, means that the lightcurve was not completely observed and that the period uncertainty can arrive up to 30%. The same U value is also used in the cases in which the true period can be a whole multiple of the observed lightcurve.

The data for the MBAs sample are drawn by the Asteroid LightCurve Data Base (or ALCDB, version of 2009 April 21), kept by Alan W. Harris and Brian D. Warner of Space Science Institute and Petr Pravec of the Astronomical Institute, Czech Republic (Warner et al., 2009). The dataset is available on http://www.minorplanetobserver.com/astlc/LightcurveParameters.htm.

The C and S classes asteroids can easily be distinguished in a (B-V)-(U-B) diagram, and appear as two homogeneous body population with different surface physical property (at least in first approximation). Asteroids of Tholen S class have an average geometrical albedo of 0.20 ± 0.07, while the C class have 0.06 ± 0.02 for the same quantity (Shevchenko and Lupishko, 1998). These values are practically independent from the diameter. The majority of MBAs have semi major axes within the range 2.2-3.2 AU. The S asteroids are located near the inner edge of the Main Belt, while the C are near the outer edge (Shevchenko and Lupishko, 1998). Generally speaking, C asteroids are more difficult to be observed than S ones, because are at a greater distance from Sun and show a lower albedo. The asteroids belonging to well known families were removed because they can have rotational properties specific of family formation that could contribute to the non-Maxwellian character of rotation frequency distribution (Pravec et al., 2002). For a similar reason, binary asteroids were removed from the list.

# 2 Description of the MBAs sample

For asteroids extraction it was written and used proper software that read the ALCDB file and save in a new file the results. The list so obtained includes 962 MBAs (495 S and 467 C). The dwarf planet (1) Ceres has been excluded because it is a peculiar object. The quantity which presents the bigger uncertainty is the diameter because it is an indirect data, obtained by calculation and/or assumption, see ALCDB readme for more information about this topic. Instead of the asteroid rotation period P (usually synodic), the rotation frequency $\Omega$ is used. The $\Omega$ value is equal to 24/P (where P is in hours), and is measured in rotations $day^{-1}$ (or rot $day^{-1}$). This quantity is directly proportional to the asteroid angular speed and angular momentum modulus. The angular momentum derives from the asteroid formation and it is modified from the subsequently collisional evolution (Harris and Burns, 1979). For small asteroids there are additional factors acting on spin such as the YORP effect which tends to flatten the rotation frequencies distribution (Rubincam, 2000, Pravec et al., 2008). The lightcurve amplitude was randomized with the Binzel and Sauter (1992) method, see Appendix for more details.

The main statistical values for C and S asteroid sample are shown in Table 1. In this Table are compared asteroids samples with the inferior diameter value gradually decreasing. As can be seen, for diameter values equal to or greater than about 20 km, the diameters distribution of C and S asteroids are about the same because the mean values and standard deviations are similar. Also, the ratio between C and S numbers is almost constant, with values between 3 and 2.7. On the contrary, below 20 km the average diameter of S is much smaller than that of C and the ratio C/S drop near 1. This difference can be attributed to an observative bias in favour of the small S asteroids.

So, the properties of the two asteroids samples are directly comparable only for diameters equal to or greater than about 20 km. In these dimension range there are not significant differences between the mean rotational frequency and amplitude of the two samples (see Table 1). If we fall below this limit, there is a difference between the rotation frequency. The S class asteroids tend to rotate about 20% more quickly than C class asteroids but this is due to the "bias effect" on small S asteroids. Instead, the mean lightcurve amplitude of the two classes are still comparable.

Table 2 shows the average values of some properties of asteroids with a diameter gradually increasing, so to have at least about 10 asteroids for bin. In Fig. 1 there is a plot of the mean bin diameter distribution. The enormous difference in asteroids number between C and S under 20 km is evident. The average data of Table 2 allow to compare the two asteroids samples.

## 3 The relationship between rotation frequency and diameter

In this paragraph we want to examine the correlation between the rotation frequency and the diameter for C and S samples. For a body population subject only to a collisional evolution, at equilibrium the rotation frequency is independent from the diameter (Harris, 1979). For rocky asteroids, the diameter limit beyond which this relation holds is about 4 km, this value is lightly beyond the minimum diameter considered in this analysis.

In Fig. 2 there is a comparative plot, frequency vs. diameter, of the two asteroids samples. Notice that for diameters less than 20 km, the S asteroids predominate on C and also are more low-frequency rotators, due to the observing bias. From Fig. 2 we can also see that there are two asteroids, (8722) Schirra for the S (diameter 5 km) and (3223) Forsium for the C (diameter 35.1 km), that have rotational frequencies, respectively, 10.41 and 10.24 rot day$^{-1}$. These values are close to the spin barrier of about 11-12 rot day$^{-1}$ found by Pravec et al. (2002), for asteroids with diameter less than 10 km. None, among the asteroids considered here, is closer to the spin barrier limit.

Both for C and S the rotation frequency turns out to be dependent from the diameter: if the diameter increases the frequency tends to decreases. The correlation is significant under the statistical point of view and this is true both considering asteroids as individual elements of the samples, and calculating the average values for frequency and diameter as those shown in Table 2. In the latter case, in a Log-Log scale, the correlation coefficient is –0.95 for C and –0.83 for S, with a probability under 5% to have a random relationship between the two mean quantities. The correlations persists by changing the bin width in a reasonable manner. The relationship between mean frequency and mean diameter may be well approximated with a cubic trend (see Fig. 3):

$$Log(\Omega_{mean}) = 0.26 + 1.43 \cdot Log(D_{mean}) - 1.40 \cdot (Log(D_{mean}))^2 + 0.34 \cdot (Log(D_{mean}))^3 \quad \text{(C asteroids)} \quad (1)$$

$$Log(\Omega_{mean}) = 0.20 + 1.57 \cdot Log(D_{mean}) - 1.59 \cdot (Log(D_{mean}))^2 + 0.42 \cdot (Log(D_{mean}))^3 \quad \text{(S asteroids)} \quad (2)$$

These relations are useful to get an idea of which rotational frequency can be expected according to the asteroid diameter and will be also used for the analysis of rotation frequencies distribution. Below about 10 km the two curves tend to a constant value of about 4-5 rot day$^{-1}$, while for larger objects the rotation frequency tends to decrease with a minimum around 100 km corresponding to a common value of about 2 rot day$^{-1}$. The trend of interpolating curves is similar to that reported by Pravec et al. (2002), using the asteroids contained in ALCDB of 2001 March 1, with U ≥ 2 (total of 984 objects) with the "running box" method. The increasing of mean rotation frequency as the mean size decreases is probably due to the fact that smaller asteroids, being the result of collisions, and having a lower inertia, can acquire a rotation frequency higher than the bigger fragments. Harris and Burns (1979), analysing 182 C and S asteroids, found out a similar trend, but without a significant correlation. Instead, a significant correlation was found by Binzel (1984), analysing MBAs with diameter equal or lower than 120 km.

## 4 The relationship between the lightcurve amplitude and the diameter

A significant correlation may also be found between the lightcurve amplitude and the diameter. For C and S MBAs, if we increase the diameter the lightcurve amplitude tend to decreases. So, as expected, the larger asteroids have a more sphere-like shape than small asteroids. The plot for the whole C/S sample is shown in Fig. 4, here the bias below 20 km is also evident, with the prevalence of S on C. For the mean values of Table 2, the correlation coefficient among the mean lightcurve amplitude and the mean diameter (in a semi-logarithmic scale), is about -0.96 both for C and S, with a probability under 5% to have a casual relationship. The relation between mean amplitude and mean diameter can be well approximated by a parabola (see Fig. 5):

$$A_{mean} = 0.40 + 0.08 \cdot (Log(D_{mean})) - 0.09 \cdot (Log(D_{mean}))^2 \qquad \text{(C asteroids)} \qquad (3)$$

$$A_{mean} = 0.45 - 0.04 \cdot (Log(D_{mean})) - 0.05 \cdot (Log(D_{mean}))^2 \qquad \text{(S asteroids)} \qquad (4)$$

The two curves tend to be parallel for large diameters but tend to converge to a constant value for small ones. Notice that, for diameters greater than 20 km where the observing bias ceases, C asteroids tend to have almost always a greater amplitude of about 0.1 magnitude with respect to S asteroids of the same diameter. However, the distance between the two curves is of the same order of the standard deviation, so that the difference is not very important. As for rotation frequency and diameters, a similar trend between amplitude and diameter, with a significant correlation, was found by Binzel (1984), for MBAs with diameter equal or lower than 120 km. The amplitude value of about 0.2 mag for the major asteroids is in agreement with rock fragments obtained in laboratory and observed from random directions (Hartmann and Tholen, 1990). Probably, the discrepancy with little asteroids is due to the different breaking way of asteroids and little stones.

## 5 Rotation frequency distribution

In a body system evolved by collisions only, one expects the spin distribution to be perfectly random in space. In this case, the three components of the angular speed vector must have a Gaussian distribution with mean value zero and identical dispersion. If these conditions are satisfied, the angular speed or the rotation frequency distribution must be a Maxwellian (Harris and Burns, 1979). The explicit form of the three-dimensional Maxwellian distribution adopted here is given in Farinella et al. (1981), or in Pravec et al. (2002).

To check the compatibility between the observed distribution and the theoretical one we have chosen to work on cumulative distributions with, at least, 100 asteroids. This limit is important to have a reasonable number of asteroids on which to apply statistical analysis. The cumulative distribution presents the asteroids number with value of the independent variable (in this case the rotation frequency), under a certain threshold. To check the compatibility with the theoretical distributions, we can use the Kolmogorov-Smirnov (K-S) or the Kuiper test, a variant of the K-S, (Press et al., 1992). The advantage of using the K-S/Kuiper test with respect to the most classical chi-square test, is that the results are independent from the bin assumed for the distribution. In our case the Kuiper test was used, because it is more sensitive than K-S to the distribution edge (Press et al., 1992). The tests were done normalizing every rotation frequency of the samples to the mean rotation frequency given by Equations (1) and (2), in their range of validity (about 4-200 km), in order to eliminate the unhomogeneity due to different asteroids sizes (Pravec and Harris, 2000). The compatibility test is considered passed if the confidence level of rejection is equal to or less than 95%.

If we consider the whole asteroids sample, C or S, the frequency distribution it's not consistent with Maxwellian/uniform distribution. As much as concerns the deviation from the Maxwellian, for both samples there is a slow rotator excess under 2 rot day$^{-1}$, a marked deficiency of rotator about 2-5 rot day$^{-1}$ and a small secondary peak at about 8 rot day$^{-1}$. The fast rotator peak of about 8 rot day$^{-1}$ is more pronounced for S than for C asteroids.

For C asteroids, if we change the inferior diameter limit from 4 to 40 km, the frequency distribution is always non-Maxwellian/non-uniform, mainly due to the persistence of slow rotators excess under 2 rot day$^{-1}$ and a rotator deficiency about 2-4 rot day$^{-1}$. However, if we take a minimum diameter of 48 km (remember that the superior limit is 200 km), the distribution is Maxwellian with a 8% compatibility (see Fig. 6). For S asteroids the behaviour is slightly different. If we change the inferior diameter limit, from 4 to 30 km the frequency distribution is always non-Maxwellian/non-uniform, always for the persistence of slow rotators excess under 2 rot day$^{-1}$ and a rotator deficiency about 2-4 rot day$^{-1}$. Only for diameters greater than or equal to 33 km the fit with the Maxwellian distribution shows a 7% compatibility (see Fig. 7). Instead, if the superior diameter limit is lowered, from 200 km to 15 km, the distribution is always non-Maxwellian/non-uniform. For diameters less than or equal to 15 km, the distribution begins to approach to the uniform one, as expected because of YORP effect. What prevents a good fit is a marked rotator deficiency between 6-7 and 9-10 rot day$^{-1}$ (see Fig. 8). Pravec et al. (2008), on MBAs/Mars crossing sample of 268 objects, with diameter between 3 and 15 km, find a uniform distribution up to 9 rot day$^{-1}$, with a slow rotators excess for $\Omega < 1$ rot day$^{-1}$. It is possible that, in S sample, the lack of asteroids with high rotation frequency, in the range 6-10 rot day$^{-1}$, is due to an observing bias on small amplitude and so fast rotating asteroids.

This dichotomous behaviour of the rotation frequency, especially for S asteroids, with a transition to the Maxwellian for larger diameters, is similar to that reported by Pravec et al. (2002). In this case, the deviations from the Maxwellian distribution are important for diameter under 30-40 km, while for objects with diameter higher than 40 km the rotation frequency distribution is compatible with a Maxwellian. Binzel (1984), for MBAs with diameter equal or lower than 30 km, finds slow rotators excess, whose existence was first noticed by Farinella et al., 1981, on a sample of 253 MBAs. For Binzel, the K-S test allows to reject the Maxwellian to 99% confidence level.

## 6 Conclusions

Let's make a summary of this comparative analysis between the C and S asteroids. Above 20 km, the diameters distributions of C and S are similar, as the average rotation frequency. Under 20 km there is an observative bias in favour of small S asteroids which prevents a comparison. To bridge this gap, photometric observations of C asteroids with a diameter under 20 km will be useful.

There is a significant correlation between rotation frequency and diameter both for C and S: if the diameter decreases the rotations frequency tends to increase. Only two asteroids in the sample (8722 and 3223) are close to the value of the spin barrier of about 11-12 rot day$^{-1}$. There is also a significant correlation between the lightcurve amplitude and the diameter for both samples: if the diameter decreases the lightcurve amplitude tends to increase. The C amplitude, for diameters greater than 20 km, tends to be systematically higher than S amplitude of about 0.1 magnitude, but the difference is not very significant.

Finally, between 48 and 200 km, the C asteroids have a rotation frequency distribution compatible with a Maxwellian while, for S asteroids, the compatibility with the Maxwellian is for diameters greater than 33 km. Below 15 km the S tend to be compatible with a uniform distribution, while for C asteroids we can not say anything because the objects with small diameters and known period are still too few.

Considering the rotational properties and the lightcurve amplitude, apart the observational bias, there appears to be no substantial differences between the samples of C and S asteroids taken into account and this indicates a good homogeneity in the collisional evolution of the two samples.

**Acknowledgments**

Research on asteroids at OAVdA Observatory is strongly supported by Director, Enzo Bertolini, and funded with a European Social Fund grant from the Regional Government of the Regione Autonoma della Valle d'Aosta (Italy). A special thanks to Peter Pravec (Astronomical Institute, Czech Republic) for the constructive comments to the manuscript, and thanks to an anonymous referee that with his advice have contributed to the improvement of the work.

**Appendix**

**The processing of the lightcurve amplitude**

This appendix describes the processing performed on the lightcurves amplitudes of the samples. The purpose of the amplitude analysis is to get information about the asteroids physical shape. Unfortunately, the spin direction of asteroids is well known for few cases only so, from a single lightcurve, it is not possible to have full shape information. To avoid this problem it is sufficient to make a statistical analysis of the well-known lightcurve amplitude for the whole asteroids sample. However, often the used samples are not homogeneous: for a few asteroids the spin and shape are well known, while for others the lightcurve only is known. Therefore the problem arises to make these data comparable. Assuming that the spin vectors have a random orientation in space, or better, that there are not preferential directions along which they are aligned, on average the observation of a single asteroid takes place with a 60° angle between the spin vector and the sight line (Binzel et al., 1989). Taking this into account, the lightcurve amplitude of the asteroids having well known pole can be "randomized" and joined to the amplitude of the ones observed only few times. The hypothesis about a random orientation of spin vectors might not be true, especially for smaller asteroids, since the YORP effect could orient it as in the case of Koronis family (Slivan, 2002). Here we have assumed a random distribution. So, to make homogeneous the lightcurve amplitudes, the algorithm of Binzel and Sauter (1992) was adopted:

1. If the asteroid was observed only once is considered the observed lightcurve amplitude.
2. If the asteroid was observed several times, are well known the minimum ($A_{min}$), and the maximum ($A_{max}$), amplitudes. If ($A_{max} - A_{min}$) ≤ 0.2 probably the asteroid was not quite sampled, so we take the average amplitude as representative, ($A_{max} + A_{min}$)/2.
3. Same as 2 but with ($A_{max} - A_{min}$) > 0.2. We consider the asteroid as well sampled, so $A_{max}$ = 2.5 log(a/b). With the a/b ratio we can obtain the amplitude for the aspect angle at 60°.
4. If the pole position and the shape of the asteroid are already well known, his amplitude is calculated for the aspect angle at 60°.

The database used for spins and asteroids shape is that of Kryszczynska et al. (2007), version of 2008 April 15. The dataset is available on http://vesta.astro.amu.edu.pl/Science/Asteroids/.

Table 1

Statistical values for sample progressively larger of C and S MBAs. The $\Omega_{mean}$ and $A_{mean}$ uncertainties are the standard deviation of the mean. The diameters distribution of the two samples, indicated by $D_{mean}$ and $\sigma_D$, are similar up to 20 km in diameter. Below this value there is an observative bias in favour of S asteroids.

| Quantity | C | S |
|---|---|---|
| **D ≥ 50 km** | | |
| Number | 170 | 56 |
| $D_{mean}$ (km) | 91.9 | 95.5 |
| $\sigma_D$ (km) | 48 | 41.6 |
| $\Omega_{mean}$ (rot day$^{-1}$) | 2.06 ± 0.08 | 2.24 ± 0.14 |
| $A_{mean}$ (mag) | 0.22 ± 0.01 | 0.193 ± 0.014 |
| **D ≥ 40 km** | | |
| Number | 222 | 82 |
| $D_{mean}$ (km) | 80.8 | 79.5 |
| $\sigma_D$ (km) | 46.5 | 41.7 |
| $\Omega_{mean}$ (rot day$^{-1}$) | 2.11 ± 0.08 | 2.23 ± 0.10 |
| $A_{mean}$ (mag) | 0.25 ± 0.01 | 0.21 ± 0.01 |
| **D ≥ 30 km** | | |
| Number | 310 | 112 |
| $D_{mean}$ (km) | 67.8 | 67.5 |
| $\sigma_D$ (km) | 44.5 | 40.8 |
| $\Omega_{mean}$ (rot day$^{-1}$) | 2.27 ± 0.08 | 2.25 ± 0.10 |
| $A_{mean}$ (mag) | 0.27 ± 0.01 | 0.23 ± 0.01 |
| **D ≥ 20 km** | | |
| Number | 398 | 148 |
| $D_{mean}$ (km) | 58.3 | 57.3 |
| $\sigma_D$ (km) | 43.1 | 39.8 |
| $\Omega_{mean}$ (rot day$^{-1}$) | 2.45 ± 0.07 | 2.29 ± 0.10 |
| $A_{mean}$ (mag) | 0.29 ± 0.09 | 0.24 ± 0.01 |
| **D ≥ 10 km** | | |
| Number | 437 | 287 |
| $D_{mean}$ (km) | 54.5 | 36.4 |
| $\sigma_D$ (km) | 42.9 | 35.9 |
| $\Omega_{mean}$ (rot day$^{-1}$) | 2.57 ± 0.08 | 3.02 ± 0.13 |
| $A_{mean}$ (mag) | 0.30 ± 0.01 | 0.30 ± 0.01 |
| **D ≥ 5 km** | | |
| Number | 459 | 442 |
| $D_{mean}$ (km) | 52.2 | 26.2 |
| $\sigma_D$ (km) | 43.1 | 32 |
| $\Omega_{mean}$ (rot day$^{-1}$) | 2.67 ± 0.08 | 3.39 ± 0.10 |
| $A_{mean}$ (mag) | 0.30 ± 0.01 | 0.31 ± 0.01 |

Table 2

Numerical mean values for diameter ($D_{mean}$, in km), rotation frequency ($\Omega_{mean}$, in rot day$^{-1}$) and amplitude ($A_{mean}$, in magnitude) vs. diameter for the C and S asteroids bin samples. The diameters range a-b mean $a \leq D < b$.

| Diam. range (km) | 1-5 | 5-10 | 10-20 | 20-30 | 30-40 | 40-50 | 50-70 | 70-90 | 90-120 | 120-500 |
|---|---|---|---|---|---|---|---|---|---|---|
| **C** | | | | | | | | | | |
| $N_C$ | 8 | 22 | 39 | 88 | 88 | 52 | 66 | 42 | 33 | 29 |
| $D_{mean}$ | 3.8 ± 0.3 | 7.2 ± 0.2 | 15.3 ± 0.5 | 25.1 ± 0.3 | 34.9 ± 0.3 | 44.5 ± 0.4 | 58.9 ± 0.7 | 78.6 ± 0.8 | 103.6 ± 1.5 | 173 ± 11 |
| $\Omega_{mean}$ | 5.00 ± 0.6 | 4.60 ± 0.5 | 3.86 ± 0.3 | 3.06 ± 0.2 | 2.67 ± 0.2 | 2.30 ± 0.2 | 1.97 ± 0.1 | 2.18 ± 0.15 | 1.97 ± 0.16 | 2.16 ± 0.24 |
| $A_{mean}$ | 0.41 ± 0.07 | 0.42 ± 0.06 | 0.38 ± 0.04 | 0.35 ± 0.02 | 0.33 ± 0.02 | 0.32 ± 0.03 | 0.27 ± 0.02 | 0.21 ± 0.02 | 0.20 ± 0.02 | 0.17 ± 0.02 |
| **S** | | | | | | | | | | |
| $N_S$ | 53 | 155 | 139 | 36 | 30 | 26 | 21 | 11 | 11 | 13 |
| $D_{mean}$ | 3.4 ± 0.2 | 7.3 ± 0.1 | 14.0 ± 0.2 | 25.7 ± 0.5 | 34.7 ± 0.6 | 45.0 ± 0.6 | 60.6 ± 2 | 81 ± 1 | 104 ± 3 | 157 ± 10 |
| $\Omega_{mean}$ | 4.54 ± 0.4 | 4.08 ± 0.2 | 3.80 ± 0.2 | 2.43 ± 0.3 | 2.28 ± 0.2 | 2.22 ± 0.2 | 2.27 ± 0.2 | 2.24 ± 0.2 | 1.57 ± 0.3 | 2.75 ± 0.3 |
| $A_{mean}$ | 0.43 ± 0.04 | 0.34 ± 0.02 | 0.35 ± 0.02 | 0.29 ± 0.03 | 0.30 ± 0.02 | 0.24 ± 0.02 | 0.24 ± 0.02 | 0.20 ± 0.03 | 0.19 ± 0.04 | 0.12 ± 0.01 |

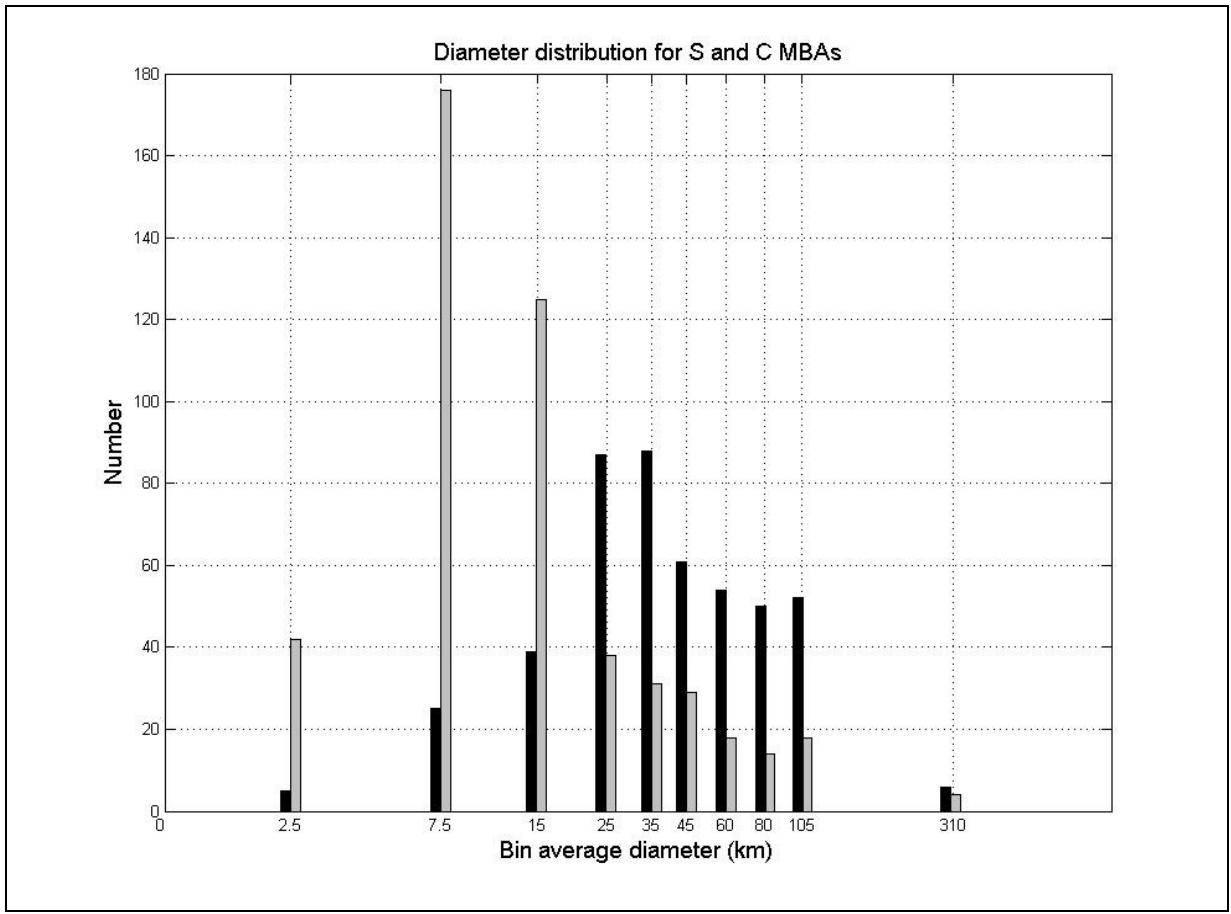

Fig. 1 Comparison between the diameter distributions for C (black bar) and S (grey bar) MBAs from data of Table 2. The two distributions are similar for diameters equal to or greater than 20 km. Below this value the amount of C asteroids collapses, while the number of S increases significantly. The diameters axis is in logarithmic scale.

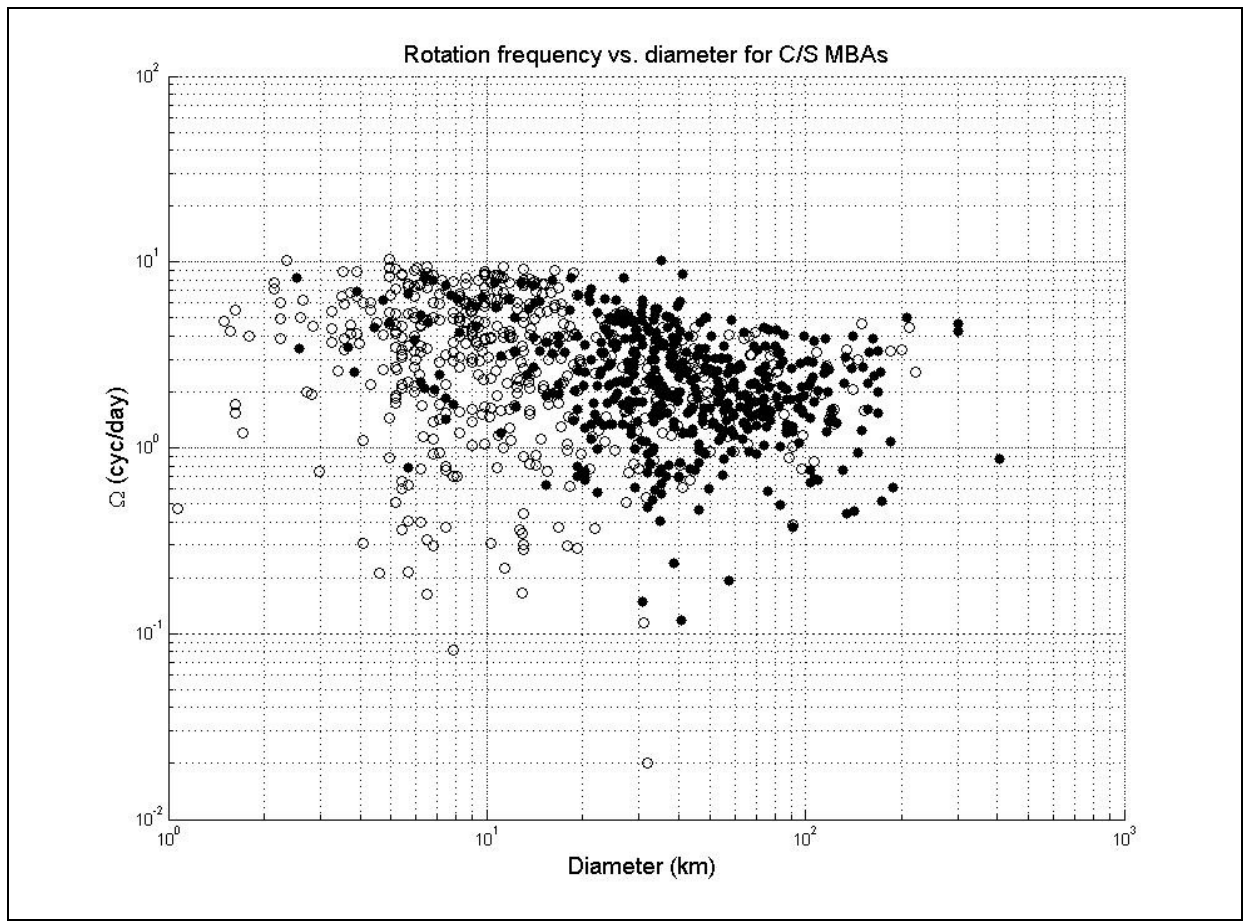

Fig. 2 Plot of the rotation frequency vs. diameter for C/S asteroids. The black dots are the C asteroids, the open circle are the S asteroids. Note that for diameters less than 20 km the S asteroids predominate on C due to the observing bias. Both axes are in logarithmic scale.

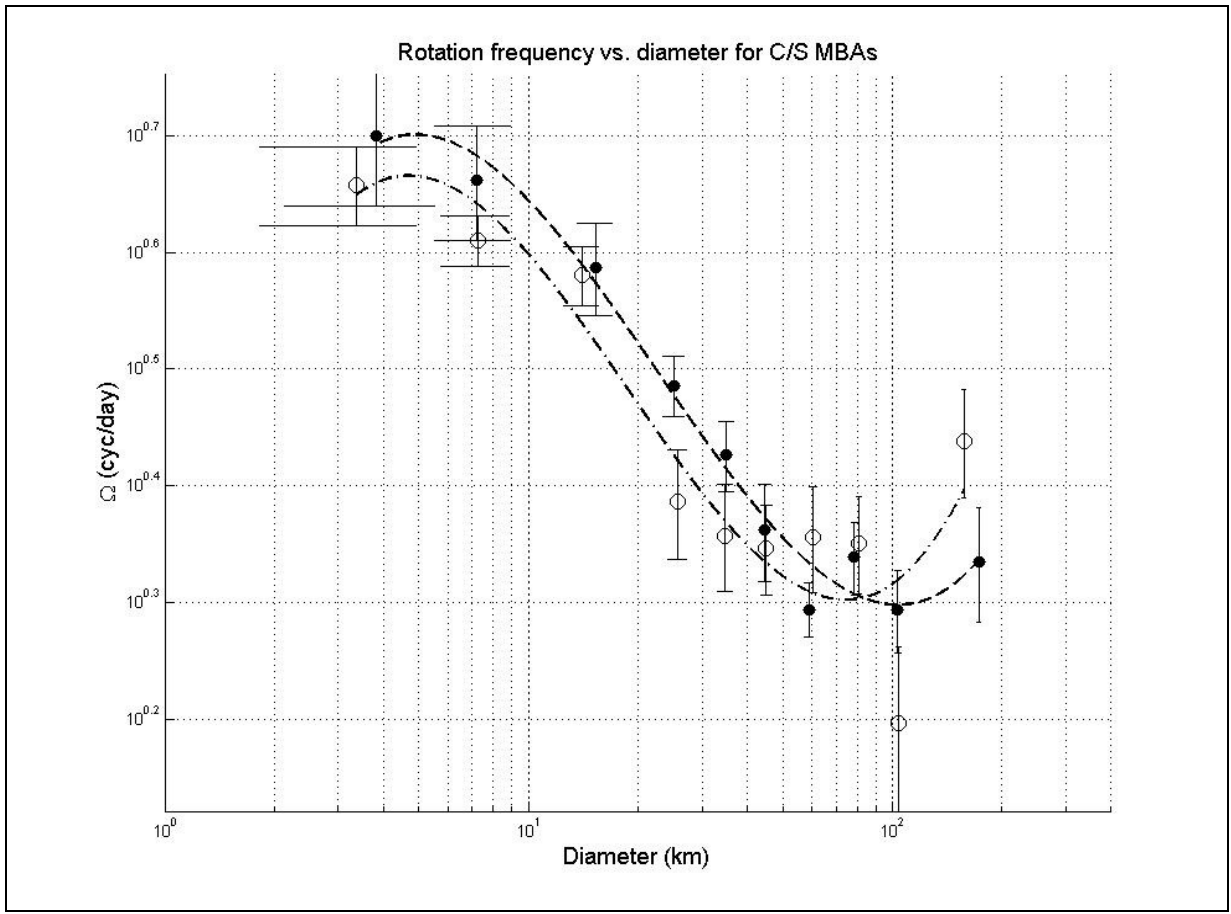

Fig. 3 The relationship among the mean rotation frequency and mean diameter from the data of Table 2. The black dots are the C asteroids, the open circle are the S asteroids. The error bars on the frequency are the standard deviation of the mean. Both axes are in logarithmic scale.

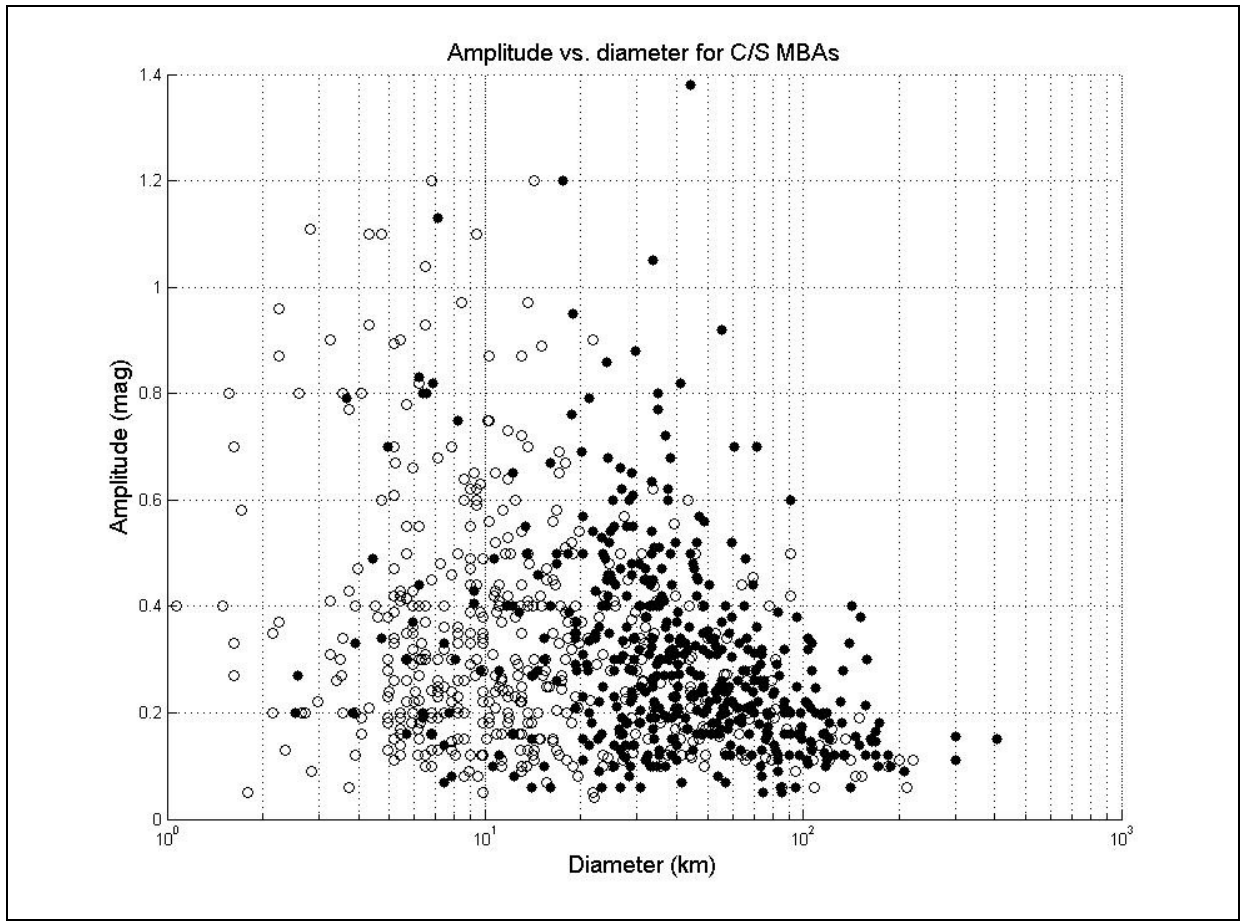

Fig. 4 Plot of the amplitude vs. diameter for C/S asteroids. The black dots are the C asteroids, the open circle are the S asteroids. Note that for diameters less than 20 km the S number predominate on C due to the observing bias. The diameters axis is in logarithmic scale.

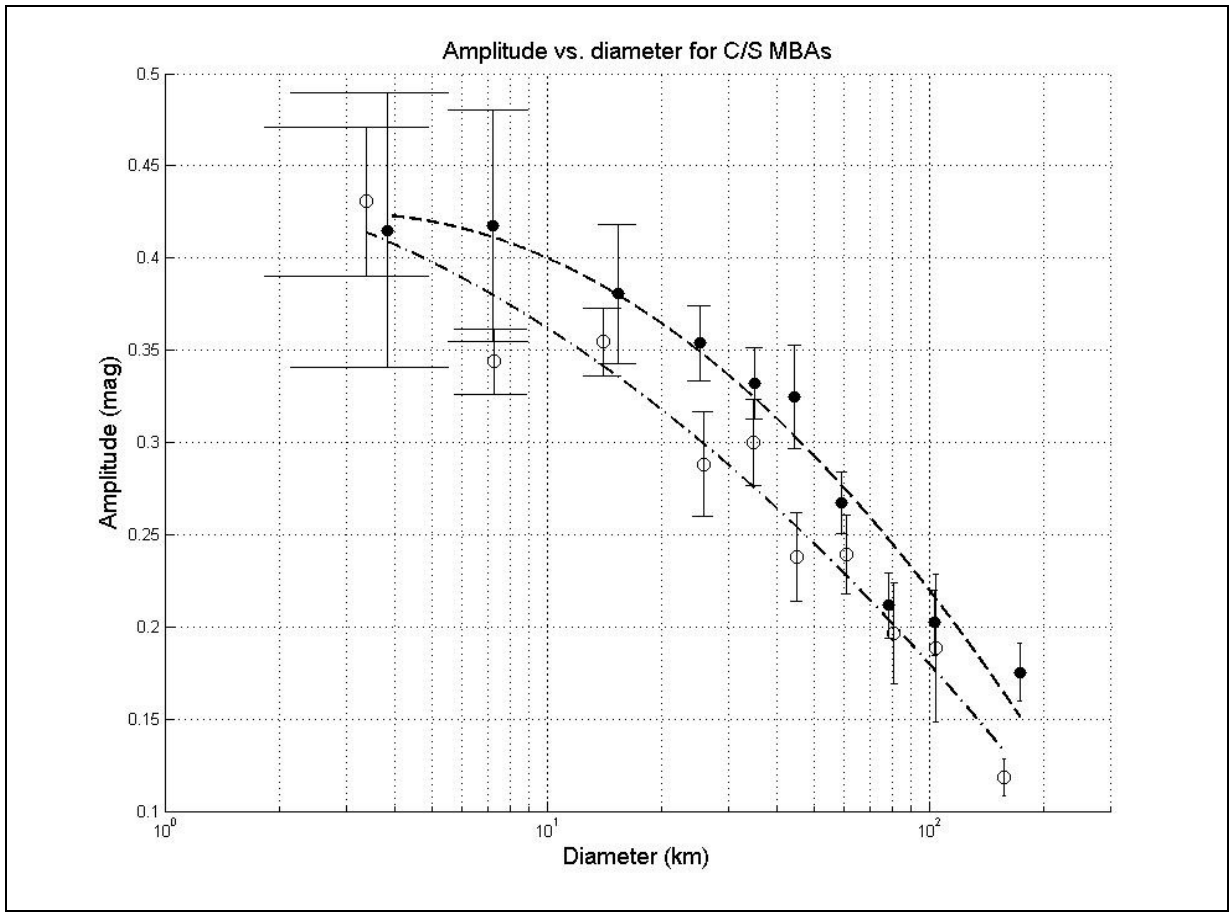

Fig. 5 The relationship among the mean amplitude and mean diameter from the data of Table 2. The black dots are the C asteroids, the open circle are the S asteroids. The error bars on the amplitude are the standard deviation of the mean. The diameters axis is in logarithmic scale.

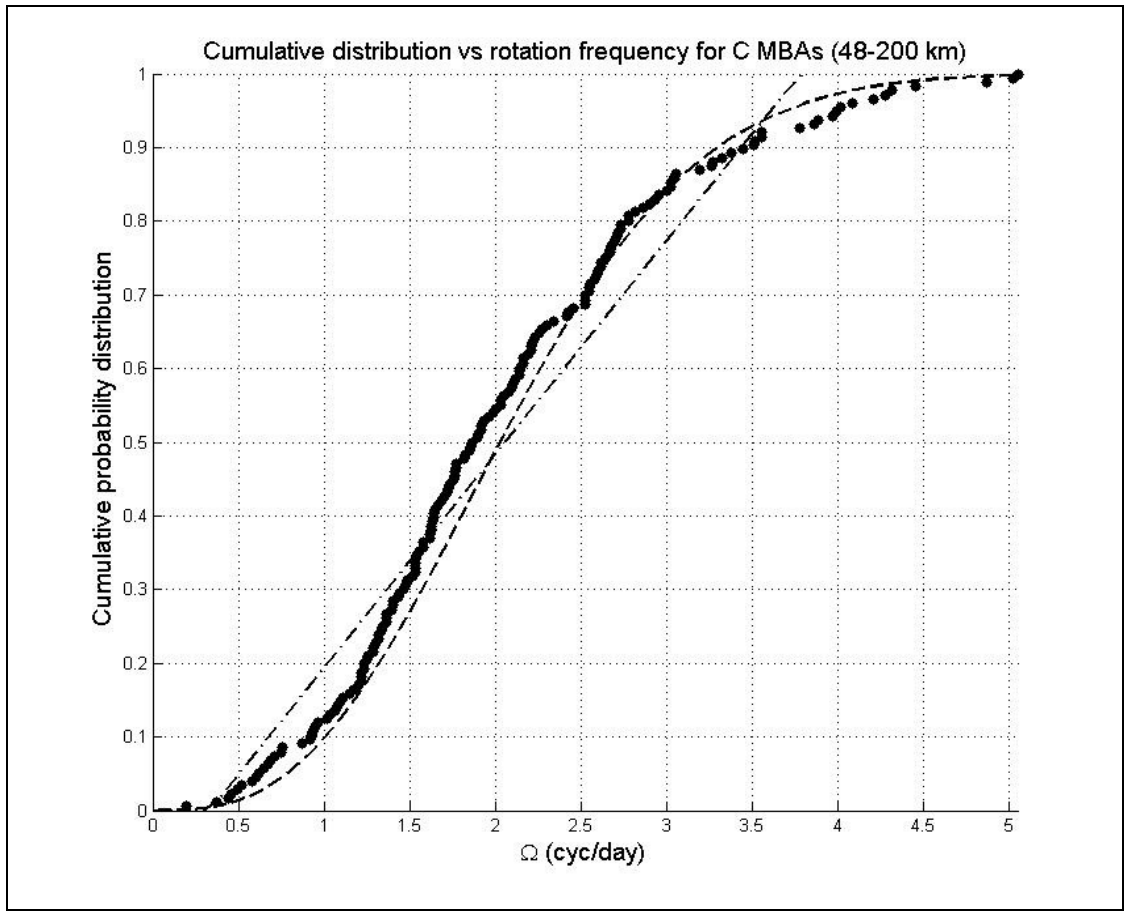

Fig. 6 Comparison among rotation frequency cumulative distributions for C MBAs with diameter from 48 to 200 km. The theoretical Maxwellian is the dashed line, the observed are the black points, and the uniform is the dashed point line.

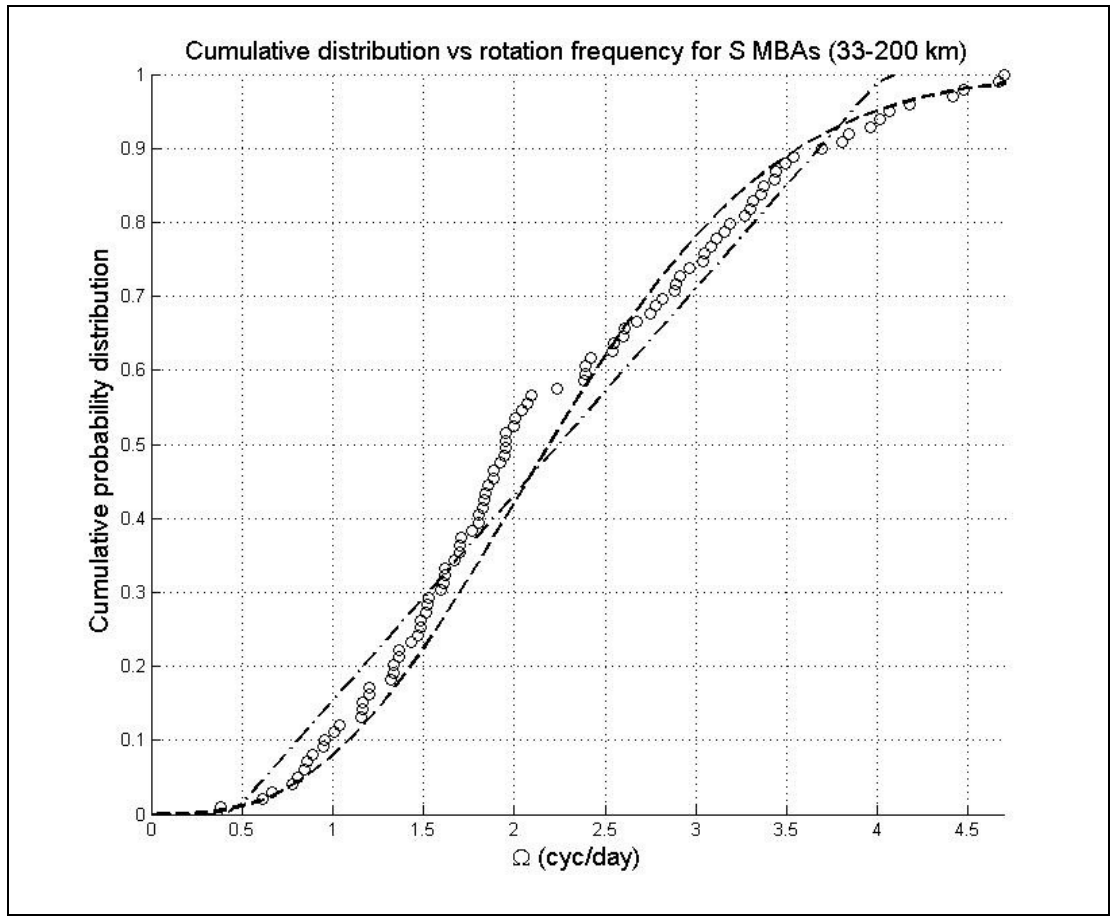

Fig. 7 Comparison among rotation frequency cumulative distributions, theoretical Maxwellian (dashed line), the observed (open circle) and the uniform (dashed point line), for S MBAs with diameter greater than or equal to 33 km.

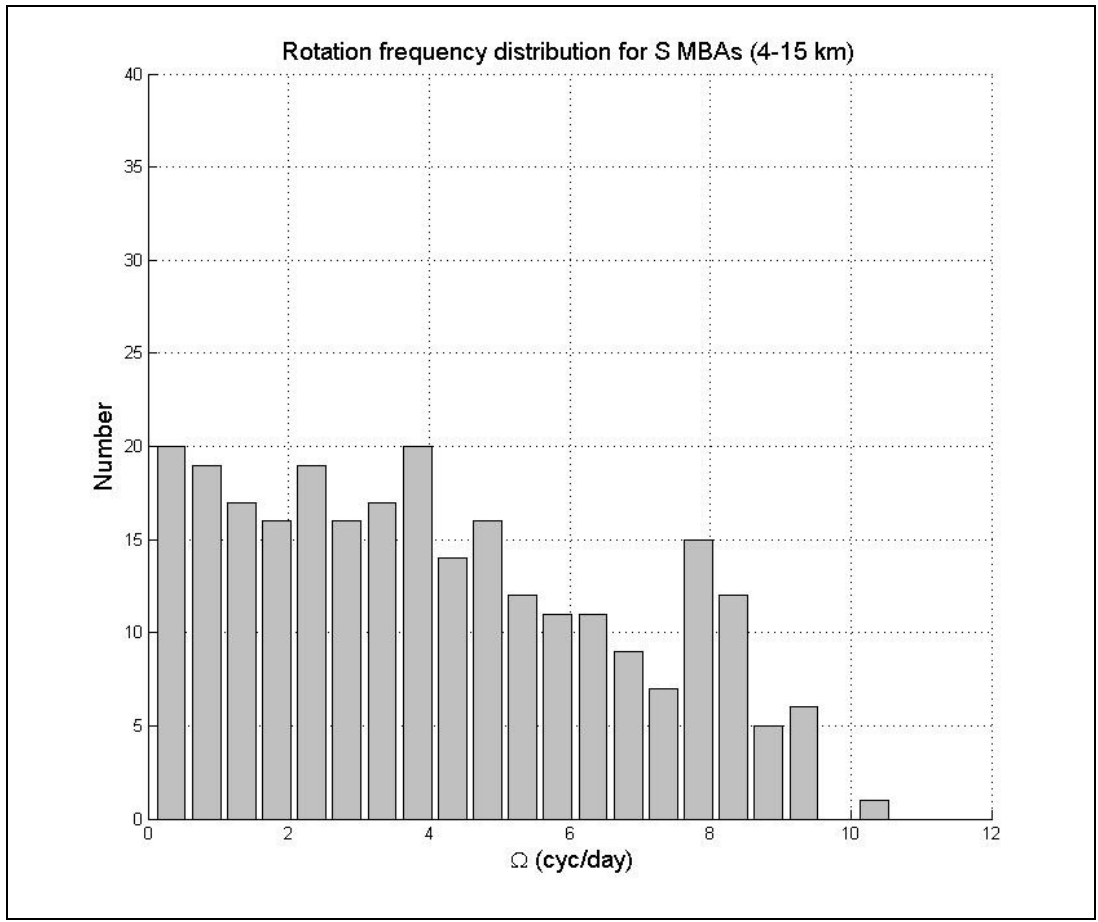

Fig. 8 Rotation frequency distributions for S asteroids with diameter in the range 4-15 km.